\documentclass[12pt]{iopart}
\usepackage{graphicx}
\usepackage{amssymb}
\usepackage{amssymb}
\usepackage{dsfont}
\usepackage{tensor}
\usepackage{upgreek}
\usepackage{braket}

\usepackage{upgreek}
\usepackage{color}
\usepackage{ulem}

\begin{document}
\pdfminorversion=4
\title[Engineering non-Markovianity from defect-phonon interactions]
{Engineering non-Markovianity from defect-phonon interactions}

\author{Francisco J. Gonz\'alez$^{1}$, Diego Tancara$^{1}$, Hossein T. Dinani$^{2}$, Raúl Coto$^{3,4}$, and Ariel Norambuena$^{\ast, 5}$}

\address{$^1$ Centro de \'Optica en Informaci\'on Cu\'antica, Camino La pir\'amide 5750, Huechuraba, Santiago, Chile}
\address{$^2$  Escuela Data Science, Facultad de Ciencias, Ingenier\'{i}a  y Tecnolog\'{i}a, Universidad Mayor, Santiago, Chile}
\address{$^3$ Department of Physics, Florida International University, Miami, Florida 33199, USA}
\address{$^4$Universidad Bernardo O~Higgins, Santiago de Chile, Chile}
\address{$^5$ Universidad Mayor, Vicerrectoría de Investigación, Santiago, Chile}
\ead{ariel.norambuena@umayor.cl}

\vspace{10pt}
%\begin{indented}
%\item[]August 2017
%\end{indented}

\begin{abstract}
Understanding defect-phonon interactions in solid-state devices is crucial for improving our current knowledge of quantum platforms. In this work, we develop first-principles calculations for a defect composed of two spin-$1/2$ particles that interact with phonon modes in a one-dimensional lattice. We follow a bottom-up approach that begins with a dipolar magnetic interaction to ultimately derive the spectral density function and time-local master equation that describes the open dynamics of the defect. We provide theoretical and numerical analysis for the non-Markovian features of the defect-phonon dynamics induced by a pure dephasing channel acting on the Bell basis. Finally, we analyze two measures of non-Markovianity based on the canonical rates and Coherence, shedding more light on the role of the spectral density function and temperature; and envisioning experimental realizations.

\end{abstract}

\vspace{2pc}
% \submitto{\NJP}
\maketitle

\section{Introduction}

Quantum Systems (QSs) are at the forefront of promising applications, including quantum computing~\cite{Kandala2017,Wright2019,Abobeih2022}, quantum machine learning~\cite{Rebentrost2014,Havlicek2019,Tancara2022,Dinani2022}, and nanoscale quantum sensors~\cite{Maze2008,Balasubramanian_2008,Coto2021}, to name a few. In general, the QS cannot be entirely shielded from the environment, and one must consider the dynamics of open quantum systems (OQSs). Modelling OQSs is a challenging task and an active field of research in condensed matter~\cite{Zhang2021,Norambuena2020}, solid-state physics~\cite{Ariel2016PRB, Caleffi2021}, and quantum optics~\cite{Xiufeng2021,Breuer2010,Orszag2016}. On the one hand, classical noise may arise from random fluctuations in an external field, such as the stochastic noise from magnetic impurities~\cite{Jamonneau2016,deLange2012,Coto2021} or the fluctuations in laser intensity~\cite{Mendoza2022}. On the other hand, quantum noise may be considered from the interaction with a fermionic~\cite{Jeske2013,Roos2020} or bosonic~\cite{Jeske2013,McCutcheon2009} bath. The latter has been crucial for explaining fundamental aspects like the radiative decay of an atom~\cite{Scully1997} or the decay of a radiation field inside a cavity QED~\cite{Scully1997}. Moreover, it also explains modern phenomena like phonon-induced relaxation in molecular defects~\cite{Cambria2021,Briganti2021}. \par

In this paper, we analyze a time-local pure dephasing dynamics of a four-level system whose environment is represented by a collection of harmonic oscillators. The four-level system is realized by two spin-$1/2$ particles, in what follows, the spin defect. The environment is considered as a thermal phononic bath in a one-dimensional lattice. Here, the key is to derive the spectral density function (SDF) that precisely describes the linear interaction between the spin defect system and the phononic bath. In particular, in solid-state devices where the quantum defect is part of the lattice, local changes in the potential energy around defect sites are created. Because of this complexity, modelling the defect-environment interaction requires a complete numerical approach. For instance, the SDF for diamond-based devices with molecular defects is generally obtained from molecular dynamics~\cite{Ariel2016PRB} or \textit{ab initio} calculations~\cite{CambriaArvix2022}. In the case of identical harmonic oscillators (without defects) with a non-uniform distribution of elastic constants a standard procedure is to calculate the SDF using the Langevin approach~\cite{Rubin1963}. Most importantly, the SDF is a valuable tool to control and understand the non-Markovian response of a quantum system~\cite{Vasile2014}. \par

Non-Markovianity is an attractive topic that, without engaging in mathematical formulations, can be recast as a scenario where the environment has short-term memory under certain interaction processes~\cite{Breuer2016,deVega2017}. Moreover, some initial quantum states can display a more sensitive response to this particular quantum memory. Non-Markovian (NM) dynamics yields more challenging descriptions of physical processes but also delivers exciting applications in state teleportation~\cite{Laine2014} and quantum metrology~\cite{Chin2012}. Therefore, engineering and understanding the environment to produce a tailored NM response through the SDF offer practical advantages. Here, we present an intuitive connection between microscopic parameters of the SDF and two measures of the degree of non-Markovianity, illustrating how a time-dependent dephasing channel acting on Bell states can generate a NM behavior. \par

The paper is organized as follows. Section~\ref{Defect-phonon} introduces the defect-phonon system, where a defect composed of two spin-$1/2$ particles is embedded in a finite-size phonon chain. Here, we present a normal mode analysis by characterizing the vibrational properties of the defect-phonon system in terms of an eigenvalue equation. In Section~\ref{spin-phonon-coupling_sec}, we derive the spin-phonon coupling from the dipolar magnetic interaction between the two spin-$1/2$ particles. These first principles calculations take into account all phonon modes of the system. In section~\ref{Spectral-Density-Function}, we explain how to obtain the spectral density function that describes the defect-phonon interaction. In particular, we present a phenomenological model that captures all physical properties, such as strong-to-weak spin-phonon coupling, broadening, and intensity. Then, in Section~\ref{Time-Local-Master-Equation}, we derive and solve the time-local master equation resulting from the spectral density function, where the dynamics is solved in the Bell basis. We conclude in Section~\ref{Non-Markovianity} by analyzing the defect's NM behavior in terms of the model's microscopic parameters and using two different NM measures.

\section{Defect-phonon system} \label{Defect-phonon}

\begin{figure}[ht!]
\centering
\includegraphics[width=0.7 \linewidth]{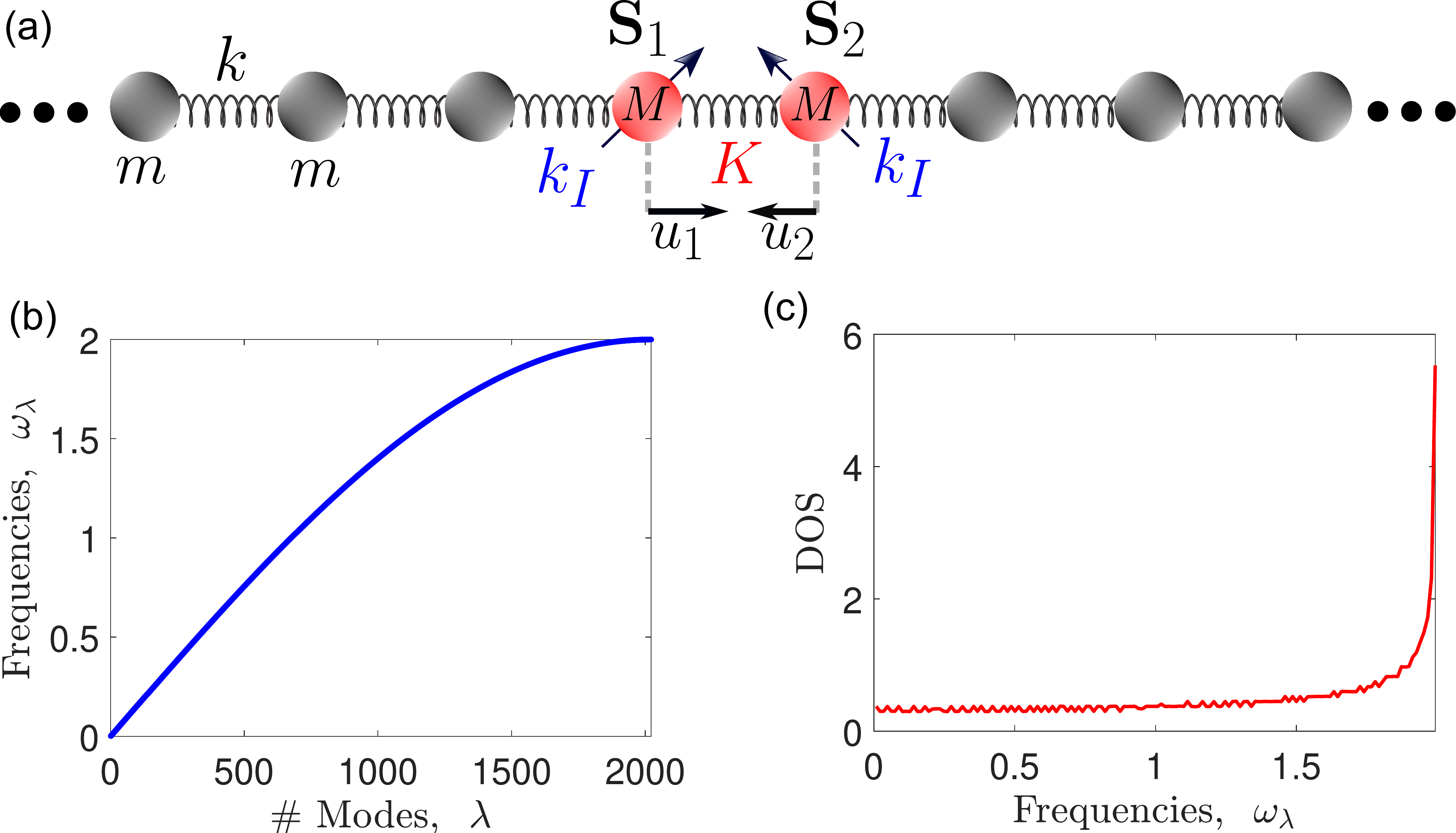}
\caption{(a) One-dimensional defect-phonon system. The system is a collection of harmonic oscillators, and the defect is composed of two spin-$1/2$ particles. (b) Phonon frequencies of the system for $N=2022$ sites, $k=K=10 k_I = 1$ and $m=M/2=1$. (c) Density of states (DOS) for the one-dimensional system.}
\label{fig:Figure1}
\end{figure}

Let us consider a one-dimensional defect-phonon system composed of a collection of $N+2$ particles, where a defect (two sites) is placed at the center of the chain ($N$ sites), as illustrated in figure~\ref{fig:Figure1}(a). The defect is composed of two spin-$1/2$ particles with masses $M$ and internal elastic constant $K$. To model vibrations in this device, we consider Newton's second law of the whole system:

\begin{equation}
M_n \ddot{u}_n = k_{n-1} (u_{n-1}-u_n)+k_{n}(u_{n+1}-u_n),  \quad n= 1-N/2,...,1,2,...,1+N/2, \label{Eq-motion}
\end{equation}

where $u_n$ are the atomic displacements around equilibrium positions, $M_n$ and $k_n$ are the mass and elastic constant through the defect-phonon system, respectively. At equilibrium ($u_n = 0$), we assume that the distance between adjacent particles is $a$ (lattice constant) such that $|u_n| \ll a$. In what follow, we use the indexes $n=1,2$ for the defect sites, and thus $M_{1,2} = M$ and $k_{1} = K$. The defect is connected with the lattice through the elastic constants $k_n = k_I$ for $n=0$ (left of $\mathbf{S}_1$) and $n=2$ (right of $\mathbf{S}_1$), see figure~\ref{fig:Figure1}(a). Elastic constants between lattice atoms with mass $m$ are considered as $k_n = k$, thus, $M_n = m$ for $n \neq 0,1$. To introduce vibrations, we apply the normal mode expansion $u_n = M_n^{-1/2}\sum_{\lambda} h_{\lambda n}Q_{\lambda}e^{i\omega t}$~\cite{Ariel2016PRB} to the Newton's second law~(\ref{Eq-motion}), yielding to the eigenvalue equation:

\begin{equation}
\sum_{j=1-N/2}^{1+N/2} D_{ij} h_{\lambda j} = \omega^2_{\lambda}h_{\lambda i}, \quad D_{ij} = {(k_i+ k_{j-1})\delta_{ij} - k_{j-1}\delta_{i,j-1}-k_{j}\delta_{i,j+1} \over \sqrt{M_i M_j}}, \label{EigenvalueEquationDij}
\end{equation}

with $D_{ij}$ representing the phononic dynamical matrix. Here, $\delta_{ij}$ is the Kronecker delta. By numerically solving the eigenvalue equation~(\ref{EigenvalueEquationDij}) one finds the phonon frequencies $\omega_{\lambda}$ and the atomic displacements $h_{\lambda i}$ for the $i$th atom in a particular mode $\lambda \in \{1,...,N+2\}$. We assume that eigenvectors are normalized such that $\sum_{i}^{}h_{\lambda i}^2 = 1$. Normal coordinates $Q_{\lambda}$ of the defect-lattice system can be computed as: 

\begin{equation} \label{NormalModesSystem}
Q_{\lambda}= \sum_{i=1-N/2}^{1+N/2}\sqrt{M_i} h_{\lambda i} u_{i},
\end{equation}

which represents collective atomic displacements for a given mode $\lambda$. Note that we have introduced the mass factor $M_i^{1/2}$ inside the normal mode expansion because of the factor $(M_i M_j)^{1/2}$ presented in the symmetrical dynamical matrix~(\ref{EigenvalueEquationDij}). In figure~\ref{fig:Figure1}(b), we show the predicted phonon frequencies $\omega_{\lambda}$ as a function of the number of modes for a system with $N=2020$. In this simulation we used the parameters $m = M/2 = 1$, $k = K = 1$, and $k_I = 0.1k$. For comparison, let us consider a one-dimensional lattice with equal masses $m$ and spring constants $k$ (without any defect) but with periodic boundary conditions. In the periodic and free-defect case, the phonon spectrum is given by $\omega_q^{\rm Per, 1D} = \sqrt{4k/m}|\sin(qa/2)|$ with $-\pi/2 \leq q \leq \pi/2$ (first Brillouin zone). When frequencies are compared, we do not observe differences between our defect-phonon system and the periodic ideal case. Moreover, in both cases, the density of phonon states (DOS) are very similar, as shown in figure~\ref{fig:Figure1}(c). However, by including a two-site defect, we shall observe that some internal vibrational modes of the defect could be strongly activated. This will be crucial to derive the phonon-induced dynamics on defect spin states, providing a route towards understanding the non-Markovian dynamics of this system. 

\section{Microscopic derivation of the spin-phonon coupling} \label{spin-phonon-coupling_sec}

In this section, we derive the spin-phonon Hamiltonian by including particle displacements as a perturbation on the dipolar magnetic interaction between two spin-$1/2$ particles. We consider that the internal dynamics of the defect is governed by the following magnetic dipolar interaction between two spin-$1/2$ particles

\begin{equation}\label{spin-spin-1}
H_{ss} = -{\mu_0 g_e^2 \mu_B^2 \over4 \pi} \left[{3(\mathbf{S}_1 \cdot \hat{r})(\mathbf{S}_2 \cdot \hat{r})-\mathbf{S}_1 \cdot \mathbf{S}_2 \over r^3}\right],
\end{equation}

where $\mu_0$ is the magnetic permeability of free space, $g_e = 2.0028$ is the Land\'e-factor of the electron, $\mu_B$ is the Bohr magneton, $\mathbf{S}_i = (1/2)(\sigma_{ix}, \sigma_{iy}, \sigma_{iz})$ are the spin operators for the $i$-th particle, $\mathbf{r} = \mathbf{r}_2-\mathbf{r}_1$, and $\hat{r} = \mathbf{r}/r$. By considering the lattice along the $x$-direction, each particle position is described by the vector $\mathbf{r}_i =  x_i \mathbf{e}_x$, where $x_i = x_i^{(0)}+u_i$, where $x_i^{(0)}$ are the equilibrium positions, and $\mathbf{e}_x = (1,0,0)^T$ is the unitary vector along $x$-direction. Then, the distance between defect sites is given by $r = |a + \delta u|$ since $x_2^{(0)}-x_1^{(0)} = a$ is the lattice constant, and $\delta u = u_2-u_1$ is a small perturbation originated from phonons. Therefore, the spin-spin Hamiltonian reduces to

\begin{equation}\label{spin-spin-2}
H_{ss} = -{C \over r^3}F(\mathbf{S}_1,\mathbf{S}_2), \quad F(\mathbf{S}_1,\mathbf{S}_2) = 2 S_{1x}S_{2x}- S_{1y}S_{2y}-S_{1z}S_{2z}
\end{equation}

with $C = \mu_0 g_e^2 \mu_B^2/(4 \pi)$. In this notation, $S_{j \alpha}$ is the $\alpha$-th component of the spin operator for the $j$-th particle. Because of the distance factor $1/r^3$ arising in the spin-spin Hamiltonian~(\ref{spin-spin-1}), we observe that the presence of phonons will perturb the internal defect dynamics. To theoretically incorporate the effect of vibrations into the spin-spin interaction, we perform the following normal mode expansion, leading to the spin-lattice Hamiltonian

\begin{equation}
H_{sl} = H_{ss}^{(0)} +  \sum_{\lambda=1}^{N+2} \left.{\partial H_{ss} \over \partial Q_{\lambda}}\right|_{u_i = 0} Q_{\lambda} + \mathcal{O}( Q_{\lambda}^2).
\end{equation}

The term $H_{ss}^{(0)} = H_{ss}(u_i=0)$ is the spin-spin interaction at equilibrium positions ($u_i=0$). The second term in the above equation is commonly called \textit{spin-lattice} Hamiltonian and it is the main ingredient to derive the open dynamics for the spin system. In this particular case, this interaction depends on the linear combination of first-order derivatives of the spin-spin interaction evaluated at equilibrium positions ($\left.\partial H_{ss} / \partial Q_{\lambda} \right|_{u_i=0}$) multiplied by the linear displacements ($Q_{\lambda}$). Quadratic terms proportional to $Q_{\lambda}^2$ in the normal mode expansion are neglected in this work but can be relevant for other solid-state systems~\cite{CambriaArvix2022}. Since the spin-spin interaction only depends on the position of defect sites, it is convenient to introduce the defect normal modes

\begin{equation} 
Q^{\rm defect}_{S} =  \sqrt{M}\left(u_1+u_2 \right), \quad Q^{\rm defect}_{A} =  \sqrt{M}\left(u_2-u_1 \right). \label{LocalNormalModes}
\end{equation}

Here, $S$ and $A$ stand for symmetric and antisymmetric normal modes, respectively. Physically, $Q_S$ ($Q_A$) represents the translational (breathing) mode of the defect. Using the defect normal coordinates, we obtain

\begin{eqnarray}
H_{sl} = \sum_{\lambda=1}^{N+2}
{\partial H_{ss} \over \partial Q_{\lambda}} Q_{\lambda} =  \sum_{\lambda = 1}^{N+2}
\sum_{\lambda'= S,A}\left( \left.{\partial H_{ss} \over \partial Q_{\lambda'}^{\rm defect}} \right|_{u_i=0}\right) \left({\partial Q_{\lambda'}^{\rm defect} \over \partial Q_{\lambda}} \right) Q_{\lambda}.
\end{eqnarray}

As the spin-spin coupling depends on the relative distance, one gets that the symmetric phonon (translational mode) does not alter this interaction, leading to $\left.\partial H_{ss} / \partial Q_{S}^{\rm defect} \right|_{u_i=0} = 0 $. On the contrary, the antisymmetrical phonon (breathing mode) changes the spin-spin interaction, and we find that $\left. \partial H_{ss} / \partial Q_{A}^{\rm defect} \right|_{u_i=0} = 3 C \sqrt{2/M} a^{-4}$. These calculations and the quantization of normal modes $Q_{\lambda} = (\hbar / 2\omega_{\lambda})^{1/2}(b_{\lambda} + b_{\lambda}^{\dagger} )$ lead to the spin-phonon Hamiltonian

\begin{equation} 
H_{\rm s-ph} = \sum_{\lambda=1}^{N} g_{\lambda} \left(b_{\lambda} + b_{\lambda}^{\dagger} \right) F(\mathbf{S}_1,\mathbf{S}_2), \label{spin-lattice}
\end{equation}

where the spin-phonon coupling constants are defined as 

\begin{equation} \label{spin-phonon-coupling}
g_{\lambda} = 3 C P_{A}(\omega_{\lambda}) \sqrt{ \hbar \over 2 M \omega_{\lambda}}a^{-4}, \quad P_{A}(\omega_{\lambda}) = \mathbf{h}_{\lambda}\cdot \mathbf{h}_{A}^{\rm defect}.
\end{equation}

Here, we detail some physical aspects of this spin-phonon interaction. First, to numerically calculate the antisymmetric projector $P_{A}(\omega_{\lambda})$ one must define the antisymmetric local coordinate of the defect ($\mathbf{h}_{A}^{\rm defect}$) using the same basis of the $N+2$ particles of the whole system, see~\ref{AppendixA} for a more detailed discussion. To this end, we define the $(N+2)$-dimensional and normalized vector $\mathbf{h}_{A}^{\rm defect}= [0,...,0,1,-1,0,...0]^T/\sqrt{2}$ which only contains defect displacements ($+1$ and $-1$ for sites $i=1$ and $i=2$, respectively). The vector $\mathbf{h}_{\lambda} = \sum_{i}^{}h_{\lambda i}\mathbf{e}_i$ (with $\mathbf{e}_i \cdot \mathbf{e}_j = \delta_{ij}$) is obtained by solving the eigenvalue equation~(\ref{EigenvalueEquationDij}), which allows to find the eigenvectors $h_{\lambda i}$. Second, the antisymmetric projection factor $P_{A}(\omega_{\lambda})$ also depends on the frequency of the mode $\omega_{\lambda}$, and for low-energy phonon modes ($\omega_{\lambda} \approx 0$) one expects that $\lim_{\omega \rightarrow 0} \omega_{\lambda}^{-1/2} P_{A}(\omega_{\lambda}) \rightarrow 0$ (including the factor $\omega_{\lambda}^{-1/2}$ arising from quantization). The latter means that the spin-phonon couplings tend to zero at low-frequency modes (translational modes of the system). Finally, the defect mass $M$, lattice constant $a$, and the factor $C$ only modify the intensity of the coupling, not the frequency distribution.

\section{Spectral density function} \label{Spectral-Density-Function}

In this section, we explore the properties of the spectral density of the system arising from the spin-phonon coupling constants $g_{\lambda}$ introduced in equation~(\ref{spin-phonon-coupling}). When a quantum system interacts with a bosonic environment, it is convenient to introduce the spectral density function (SDF) that captures the information about the system-environment coupling. From the spin-phonon Hamiltonian~(\ref{spin-lattice}), the SDF is defined as

\begin{equation} \label{NumericalSDF}
J(\omega) = \sum_{\lambda}|g_{\lambda}|^2 \delta(\omega-\omega_{\lambda}),  \quad 0 \leq \omega \leq \omega_{\rm max},
\end{equation}

\begin{figure}[ht!]
\centering
\includegraphics[width=1\linewidth]{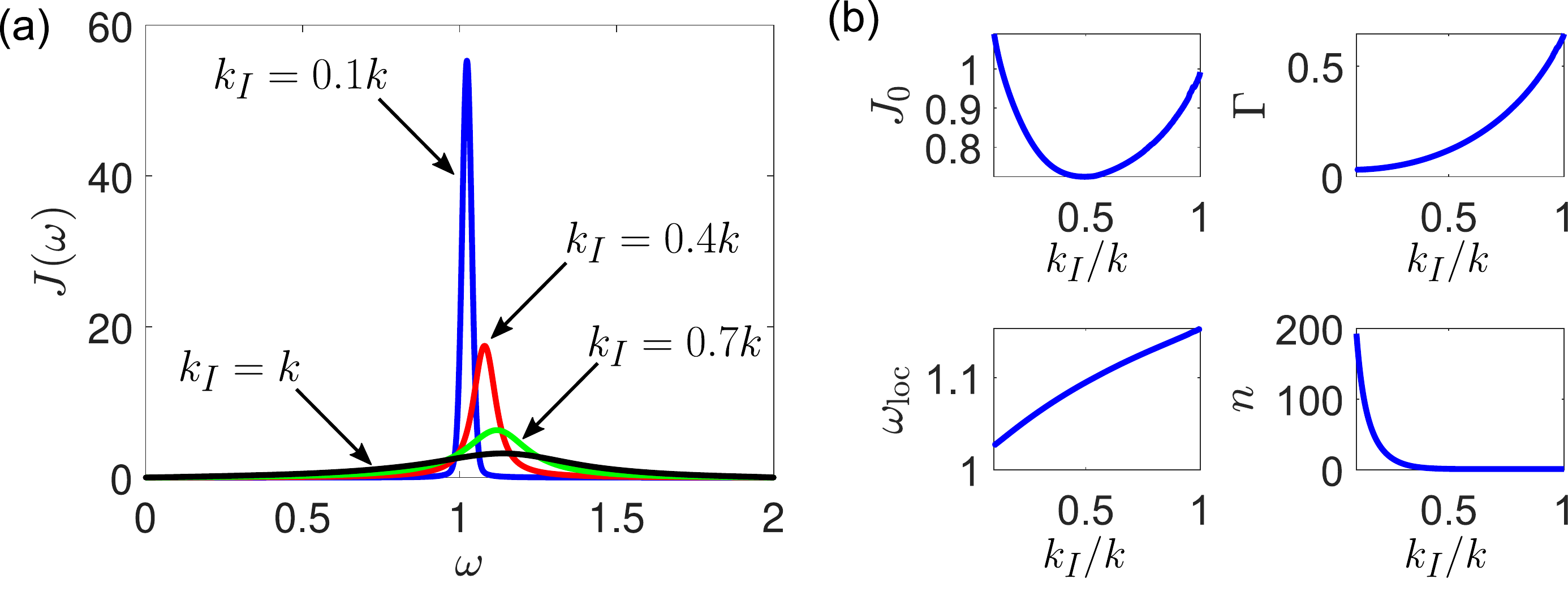}
\caption{(a) Numerical spectral density function for different defect-lattice interactions $k_I$. (b) Fit parameters of the phenomenological spectral density function given in Eq.~(\ref{SDF}) as a function of $k_I/k$.}
\label{fig:Figure5}
\end{figure}

where $|g_{\lambda}|^2$ is known as the \textit{partial Huang-Rhys (HR) factor}~\cite{HRfactor} and $\delta(x)$ is the Dirac delta function. In general terms, the SDF $J(\omega)$ satisfies two physical properties: i) $J(0) = J(\omega>\omega_{\rm max})=0$ and ii) $J(\omega) > 0 \; \forall \; \omega \in (0, \omega_{\rm max})$, where $\omega_{\rm max} = \sqrt{4k/m}$ is the largest frequency of the system. As a consequence, the SDF satisfies the integral condition

\begin{equation} \label{ConstraintJ}
    \int_{0}^{\omega_{\rm max}}J(\omega)\, d\omega = \sum_{\lambda}|g_{\lambda}|^2,  
\end{equation}

and therefore, the intensity of the SDF is constrained by the sum of partial HR factors. To find a continuous expression for $J(\omega)$, we use the Gaussian representation for the Dirac delta function $\delta(\omega-\omega_{\lambda})$, which allows us to write the following expression (Gaussian smearing):

\begin{equation}
J(\omega) = \lim_{\sigma \rightarrow 0} \sum_{\lambda} {1 \over \sqrt{\pi} |\sigma|}|g_{\lambda}|^2 e^{-(\omega-\omega_{\lambda})/\sigma^2}.
\end{equation}

From a numerical point of view, we note that a width $\sigma = 10^{-2}\omega_{\rm max}$ is enough to capture the smooth behavior of $J(\omega)$. In figure~\ref{fig:Figure5}(a), we plot different SDFs by changing the elastic constant $k_I$ that connects the defect to the phonon environment. The Lorentzian shape of $J(\omega)$ motivates us to find a phenomenological model for further applications in open dynamics. Therefore, we introduce the following phenomenological SDF:

\begin{equation}
J(\omega) = J_0 \sin^{1+n}\left({\omega \over \omega_{\rm max}}\pi \right){\Gamma/2 \over  \left[\left(\omega - \omega_{\rm loc} \right)^2 +(\Gamma/2)^2 \right]}. \label{SDF}
\end{equation}

The spirit of the above SDF is to capture narrow-to-broad spin-phonon coupling with the vibrational modes. We remark that, to the best of our knowledge, this phenomenological SDF is not reported in the literature, but it is inspired by a generalization of Refs.~\cite{Wilson-Rae-PRB, Ariel2016PRB}. Physically, out-of-phase oscillations of the dimer defect with large amplitude lead to a strong spin-phonon interaction (peak of $g_{\lambda}$), as shown in figure~\ref{fig:Figure5}(a). This is usually modeled by Lorentzian-like functions centered around the specific localized phonon frequency $\omega_{\rm loc}$, with a characteristic width $\Gamma$ and intensity $J_0$. When $\Gamma \rightarrow 0$ ($\Gamma \rightarrow \infty$), the SDF becomes narrow (broad) and with a high (small) amplitude at $\omega = \omega_{\rm loc}$. The term $\sin^{1+n}\left({\omega \over \omega_{\rm max}\pi} \right)$ is introduced to ensure the physical constraint $J(0) = J(\omega_{\rm max}) = 0$. The index $1+n$ ($n \geq 1$) is used to narrow the contribution of $\sin(\cdot)$, like a Gaussian effect. \par

It is important to analyze how the SDF changes for different values of the elastic constant $k_I$. Physically, $k_I$ models how rigid or soft is the restitution force resulting from defect-phonon interactions. In some cases, a force-constant model simulates the molecular bonds arising from quantum mechanical principles~\cite{Musgrave&Pople,Ariel2016PRB}. Thus, the elastic constant $k_I$ could be related, for atomic or molecular chains, with the value of the bond around the defect sites. In summary, we have four fitting parameters $(J_0, \Gamma, \omega_{\rm loc},n)$ for the SDF, whose dependency in terms of the elastic constant $k_I$ is shown in figure~\ref{fig:Figure5}(b). Using this set of fitting parameters, we obtain a chi-squared above $\approx 0.998$ compared to the numerical spectral function defined in equation~(\ref{NumericalSDF}), validating our phenomenological model. Now, we have all ingredients to model the non-Markovian dynamics of the defect. \par

\section{Time-local pure-dephasing dynamics} \label{Time-Local-Master-Equation}

In this section, we derive the open dynamics of the defect arising from the spin-phonon coupling presented in the previous section. The total Hamiltonian of the system is given by ($\hbar = 1$)

\begin{equation}
  H = H_{\rm s}+H_{\rm ph}+H_{\rm s-ph}, 
\end{equation}

where $H_{\rm s}  = C/a^3 F(\mathbf{S}_1,\mathbf{S}_2)$ is the defect Hamiltonian (at equilibrium positions), $H_{\rm ph} =\sum_{\lambda} \omega_{\lambda} b_{\lambda}^{\dagger}b_{\lambda}$ is the phonon Hamiltonian, and $H_{\rm s-ph} = \sum_{\lambda} g_{\lambda} (b_{\lambda} + b_{\lambda}^{\dagger} ) F(\mathbf{S}_1,\mathbf{S}_2)$ is the spin-phonon Hamiltonian. The eigenvectors of the defect Hamiltonian, \textit{i.e.} $H_{\rm s}\ket{i} = E_i \ket{i}$, are described by the following Bell-states:

\begin{eqnarray}
\ket{1}&=& \frac{1}{\sqrt{2}}\left(\ket{\uparrow\uparrow}-\ket{\downarrow\downarrow}\right), \quad E_1=E, \label{e1}\\
\ket{2}&=& \frac{1}{\sqrt{2}}\left(\ket{\uparrow\downarrow}-\ket{\downarrow\uparrow}\right), \quad E_2=0, \label{e2}\\
\ket{3}&=& \frac{1}{\sqrt{2}}\left(\ket{\uparrow\uparrow}+\ket{\downarrow\downarrow}\right), \quad E_3=-{E \over 2}, \label{e3} \\
\ket{4}&=& \frac{1}{\sqrt{2}}\left(\ket{\uparrow\downarrow}+\ket{\downarrow\uparrow}\right), \quad E_4=-{E \over 2}, \label{e4}
\end{eqnarray}

where $\ket{\alpha \beta} = \ket{\alpha}_1 \otimes \ket{\beta}_2$ are the two spin states for $\alpha,\beta = \{\uparrow, \downarrow \}$, and $E = C/a^3$ is the characteristic energy of this system. Hereafter we describe the dimer defect in this Bell basis. 

\begin{figure}[ht!]
\centering
\includegraphics[width=0.7 \linewidth]{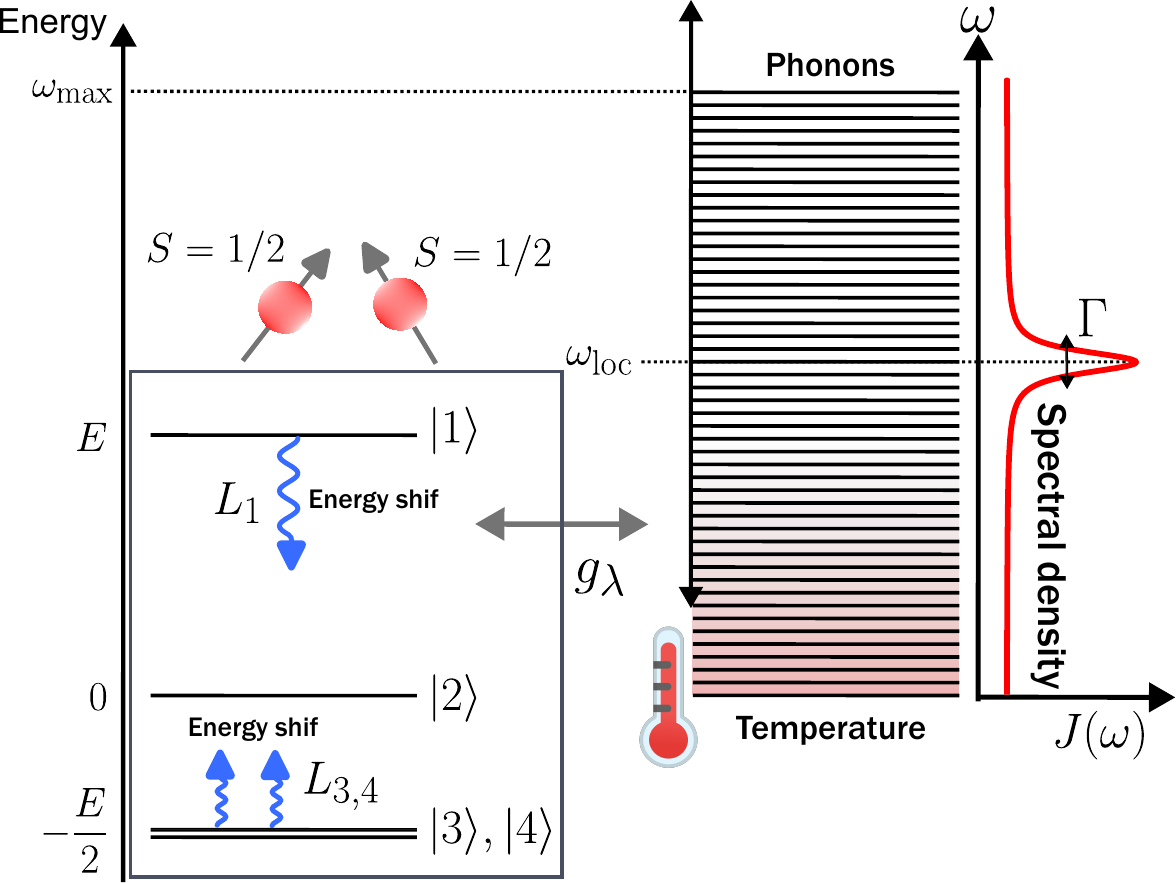}
\caption{Energy levels of the dimer defect and diagram illustrating the open dynamics induced by phonons. The two spin-$1/2$ particles originate a four-level system composed of Bell states, see equations~(\ref{e1})-(\ref{e4}), where the ground state ($|3\rangle$,$|4\rangle$) is degenerate. The spin-phonon coupling constants $g_{\lambda}$ introduce a pure dephasing effect on states $|1\rangle$, $|3\rangle$ and $|4\rangle$. The spectral density function $J(\omega)$ is modeled by a Lorentzian-like function with a peak around the localized frequency $\omega_{\rm loc}$, and a full width at half maximum (FWHM) $\Gamma$.}
\label{fig:EnergyLevels}
\end{figure}

Let us take a look at the energy spectrum of the defect system. We note that the defect exhibit a double degenerated ground state given by states $|3\rangle$ and $|4\rangle$ ($E_3 = E_4 = -E/2$). Furthermore, the state $|2\rangle$ is a dark state with zero energy ($E_2 = 0$), and the excited state $|1\rangle$ has energy $E_1 = 2|E_{3,4}|$. The state $|2\rangle$ may be relevant for quantum information processing, as it can encode decoherenceless information \cite{Gonzalez2022}. \par

Now, let us define $\rho_{\rm T}(t)$ as the total density matrix of the defect-phonon system. First, we assume the Born approximation in the Schr\"{o}dinger picture, $\rho_{\rm T}(t) =\rho_{\rm s}(t)\otimes\rho_{\rm ph}$, where $\rho_{\rm ph} = \mbox{exp}(-\beta H_{\rm ph})/Z$ is the thermal state for phonons (equilibrium bath) with $Z = \mbox{Tr}(\mbox{exp}(-\beta H_{\rm ph}))$ representing the partition function, and $\beta = (k_B T)^{-1}$ describing the inverse temperature. The state of the defect is described in terms of the reduced density matrix $\rho_{\rm s}(t) = \mbox{Tr}_{\rm ph}(\rho_{\rm T})$. In the interaction picture, and using secular and first-Markov approximation, we obtain the following time-local master equation ($\hbar = 1$): 

\begin{equation} \label{TimeLocalMasterEquation}
\dot{\rho}(t) = \gamma_c(t)\left[L \rho(t) L^{\dagger} - {1 \over 2}\{L^{\dagger} L,\rho(t) \} \right], 
\end{equation}

where $\rho(t) = U_0^{\dagger}(t) \rho_{\rm s}(t) U_0(t)$ is the reduced density matrix of the defect written in the interaction picture, where $U_0(t) = \mbox{exp}(-i H_0 t/\hbar)$ and $H_0 = H_{\rm s}+H_{\rm ph}$. In ~\ref{AppendixB}, we present a detailed derivation of the open dynamics. In the Bell basis $|i\rangle$ (see equations~(\ref{e1})-(\ref{e4})), the operator $L$ is described by a pure-dephasing channel

\begin{equation}
 L = L_1 + L_3 + L_4 =  -\sqrt{{2 \over 3}}\left[\ket{1}\bra{1}-{1 \over 2}\left(\ket{3}\bra{3}+\ket{4}\bra{4} \right) \right],  
\end{equation}

where $L_i = \langle i | L | i\rangle |i\rangle\langle i|$. Physically, the dephasing operator $L$ shifts the degenerate states $|3\rangle$ and $|4\rangle$ with the same intensity. Therefore, no coherence can be generated between these states. In addition, this dephasing channel does not alter the dark state $|2\rangle$, and the state $|1\rangle$ is shifted in the opposite direction of states $|3\rangle$ and $|4\rangle$, with double intensity. A schematic representation of the energy levels of the defect, the spin-phonon coupling, and the effect of the dephasing operator $L$ is illustrated in figure~\ref{fig:EnergyLevels}. The dephasing operator $L$ satisfies the normalization condition $\mbox{Tr}(L^{\dagger}L) = 1$, and thus, $\gamma_c(t)$ is the canonical rate~\cite{Hall}. The canonical rate $\gamma_c(t)$ is defined as

\begin{equation} \label{canonicalrate}
  \gamma_c(t)= 3\int_{0}^{\infty} {J(\omega) \over \omega} \mbox{coth}\left({\hbar \omega \over 2 k_B T} \right) \sin(\omega t) d\omega,
\end{equation}

where $J(\omega)$ is the SDF of the system~(\ref{NumericalSDF}) and $\mbox{coth}(\hbar \omega /2 k_B T) = 2n(\omega)+1$ is the effective thermal occupancy factor derived from the Bose-Einstein distribution $n(\omega) = [\mbox{exp}(\hbar \omega / k_B T) - 1]^{-1}$. In the low-temperature regime ($T \ll \omega_{\rm loc}$), we have $\mbox{coth}(\hbar \omega /2 k_B T) \approx 1$, leading to a rate $\gamma_c(t)$ with a small amplitude. Therefore, the rate amplitude increases significantly when the temperature increases. Most importantly, this canonical rate $\gamma_c(t)$ could be negative in some time intervals, depending on the nature of the SDF $J(\omega)$. The possibility of having negative values for the rate $\gamma_c(t)$ leads to a more interesting scenario than purely Markovian open quantum dynamics. Furthermore, since we have microscopically derived the SDF, we will explain some interesting non-Markovian features in connection with the spin-phonon coupling.

\section{Non-Markovianity} \label{Non-Markovianity}

\subsection{Criteria based on canonical rates}

Let us consider our first measure to quantify the degree of quantum non-Markovianity (NM). If a $d$-dimensional Lindblad master equation is written in the form $\dot{\rho}=\sum_{j=1}^{d^2-1}\gamma_j^c(t)(L_j(t)\rho(t)L_j^\dagger(t) - \lbrace L_j^\dagger(t)L_j(t),\rho(t)\rbrace /2)$~\cite{Hall} and $\mbox{Tr}( L_i^\dagger(t) L_j(t) )=\delta_{ij}$, then $\gamma_j^c(t)$ are known as the canonical rates. By a simple inspection, we observe that our time-local master equation~(\ref{TimeLocalMasterEquation}) satisfies these conditions mentioned above. Based on this idea, we use the following measure of the degree of NM~\cite{Hall,Rivas},

\begin{equation}\label{N_gamma}
	\mathcal{N}_{\gamma} = {1 \over 2}\int_{0}^{t} \left(\vert\gamma_c (\tau) \vert -\gamma_c (\tau)\right)\, d\tau.
\end{equation}

\begin{figure}[ht!]
\centering
\includegraphics[width=1\linewidth]{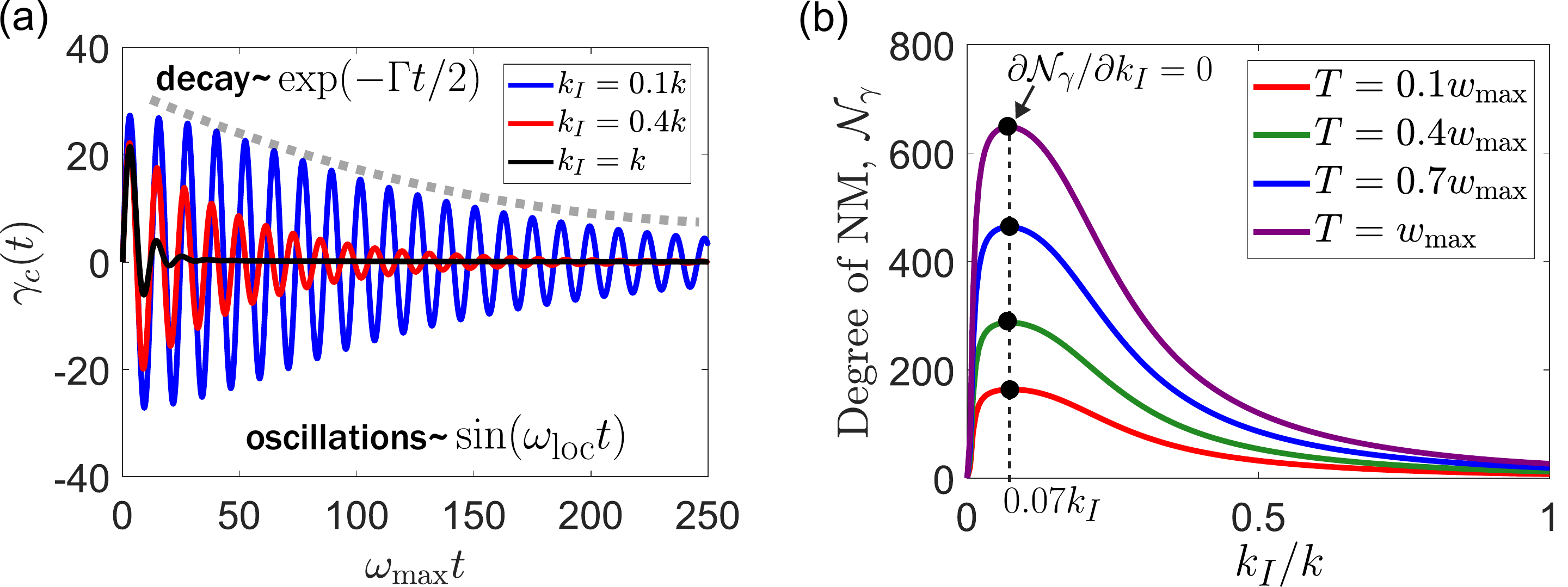}
\caption{(a) Temporal behavior of the canonical rate $\gamma_c(t)$ for different defect-lattice elastic constants $k_I$. For the simulations, we use a fixed temperature $T = \omega_{\rm max}$, and $m=M/2=1$. The exponential envelope ($\sim \mbox{exp}(-\Gamma t/2)$) and oscillations ($\sim \sin(\omega_{\rm loc}t)$) are illustrated to highlight the effect of the parameters $(\Gamma, \omega_{\rm loc})$ of the phenomenological SDF $J(\omega)$ given in equation~(\ref{SDF}). (b) Degree of NM, see equation~(\ref{NM_theo}), based on the canonical rate as a function of $k_I/k$. Different curves show the effect of temperature on the degree of NM.}
\label{fig:Figure6}
\end{figure}

The above measure of NM is only based on the accumulative negative values of the canonical rates; therefore, the dynamics is irrelevant. In figure~\ref{fig:Figure6}(a), we plot the canonical rate $\gamma_c(t)$ for three different values of the elastic constant $k_I$ at a fixed temperature $T = \omega_{\rm max}$ (maximum phonon frequency). At low temperatures, the temporal nature of $\gamma_c(t)$ is similar, but the oscillation amplitude is decreased. We note that the rate exhibits oscillations because of the strong coupling described in the spectral density function $J(\omega)$ at the frequency $\omega = \omega_{\rm loc}$, see figure~\ref{fig:Figure5}. However, the width of the spectral function ($\Gamma$) introduces an exponential envelope effect on the canonical rate (dissipative effect). 

The combined effect of $\omega_{\rm loc}$ and $\Gamma$ explains the damped oscillations observed for the canonical rates. These observations motivate the following approximation 

\begin{equation}
    \gamma_c(t) \approx \Lambda(T) \sin(\omega_{\rm loc}t)e^{-\Gamma t/2}, \quad 
    \Lambda(T) = \alpha \mbox{coth}\left({\hbar \omega_{\rm loc} \over 2 k_B T}\right),  \label{RateTheo}
\end{equation}

where $\alpha  = 3\pi \omega_{\rm loc}^{-1} J_0 \sin^{1+n}(\pi\omega_{\rm loc}/\omega_{\rm max})>0$ is a phenomenological temperature-independent factor (see ~\ref{AppendixC} for more details). The above expression is valid for values $0<k_I \leq k$. The amplitude of the canonical rate $\gamma_c(t)$ is multiplied by a thermal factor $\Lambda(T) \propto \mbox{coth}(\hbar \omega / 2 k_B T) = 2n(\omega)+1$, which explains the temperature effect observed in figure~\ref{fig:Figure6}(b). Note that even at zero temperature, we get a non-zero rate $\gamma_c(t, T=0) \approx \alpha \sin(\omega_{\rm loc}t)\mbox{exp}(-\Gamma t/2)$ since $\Lambda(T=0) = \alpha$. Using the expression~(\ref{RateTheo}) into the definition of $\mathcal{N}_{\gamma}$, we get the approximated formula in the limit $t \rightarrow \infty$ (see ~\ref{AppendixC})

\begin{equation} \label{NM_theo}
    \mathcal{N}_{\gamma} \approx {1 \over 2} \Lambda(T) \mbox{coth}\left( {\pi \Gamma \over 2\omega_{\rm loc}}\right){\omega_{\rm loc} \over \omega_{\rm loc}^2 + (\Gamma/2)^2}.
\end{equation}

In figure~\ref{fig:Figure6}(b), we observe the degree of NM according to equation~(\ref{NM_theo}), which exhibits an interesting behavior in terms of the elastic constant $k_I$. For $k_I = 0$, the spin defect is disconnected from the phonon environment, and thus the SDF $J(\omega) = 0$, leading to $\gamma_c(t) = 0$ (or equivalently $\mathcal{N}_{\gamma}=0$). At very low values of $k_I$ ($k_I/k \ll 1$), the defect particles can oscillate with a large amplitude because of the weak restoring force acting on the defect sites. Consequently, this leads to a strong and narrow spin-phonon coupling responsible for a large degree of NM (maximum value \textit{i.e.} $\partial \mathcal{N}_{\gamma}/\partial k_I  = 0$). However, when $k_I$ approaches to $k$, the SDF $J(\omega)$ is broadened at the same time that the amplitude is decreased (see figure~\ref{fig:Figure5}(a) and constraint~(\ref{ConstraintJ})), leading to a small degree of NM. Having explained the connection between the SDF $J(\omega)$ and the degree of NM $\mathcal{N}_{\gamma}$, we will now turn our attention to dynamical aspects of this system.

\subsection{Criteria based on Coherence}

For a pure-dephasing channel, one good witness for quantum NM is the Coherence~\cite{LoFranco,Rivas2017}. Also, from an experimental point of view, the Coherence will be connected to measurements, as we will detail later. Let us consider our complete set of states $\{|i\rangle\}$ given by Eqs.~(\ref{e1})-(\ref{e4}). The general solution of the time-local master equation~(\ref{TimeLocalMasterEquation}) is given by

\begin{equation}
    \rho_{ii}(t) =  \rho_{ii}(0), \quad  \rho_{ij}(t) = \rho_{ij}(0)e^{-\xi_{ij} \Upsilon(t)} \; (i\neq j), \label{generalsolutionrho}
\end{equation}

where $\rho(0) = \sum_{ij} \rho_{ij}(0)|i\rangle \langle j|$ is a generic initial condition. As expected for a pure-dephasing channel, diagonal elements (populations) are constant, and only off-diagonal elements (Coherence) change over time. The function $\Upsilon(t)$ that appears in the off-diagonal matrix elements~(\ref{generalsolutionrho}) reads

\begin{eqnarray} \label{Gamma}
  \Upsilon(t) &=& \int_{0}^{t} \gamma_c(\tau) d\tau = 3\int_{0}^{\infty} {J(\omega) \over \omega^2} \mbox{coth}\left({\hbar \omega \over 2 k_B T} \right) \left[1-\cos(\omega t) \right] d\omega.
\end{eqnarray}

We note that the function $\Upsilon(t)$ is a bounded non-negative function satisfying $0 \leq \Upsilon(t) \leq 6\int_{0}^{\infty} J(\omega)/\omega^2 \mbox{coth}(\hbar \omega / 2 k_B T) d\omega$. The parameters $\xi_{ij}$ are the elements of the matrix

\begin{equation} \label{Gamma_xi}
 \xi = \left(\begin{array}{cccc}
      0 & \xi_1 & \xi_2 & \xi_2 \\
      \xi_1 & 0 & \xi_1 & \xi_1  \\
      \xi_2 & \xi_1 & 0 & 0 \\
      \xi_2 & \xi_1 & 0 & 0 \\
  \end{array} \right), 
\end{equation}
 
 where $\xi_1 = 1/3$ and $\xi_2 = 2 \xi_1$. We remark that matrix elements $\xi_{34} = \xi_{43} = 0$ because of the degeneracy of Bell states $|3\rangle$ and $|4\rangle$. Coherence can be defined as $C(t)=\sum_{i\neq j}|\rho_{ij}(t)|$~\cite{Baumgratz}, and using the general solution~(\ref{generalsolutionrho}), we get

\begin{equation}
    C(t) = 2 \sum_{i<j} |\rho_{ij}(0)| e^{-\xi_{ij} \Upsilon(t)}. \label{Coherence}
\end{equation}

\begin{figure}[ht!]
\centering
\includegraphics[width=1\linewidth]{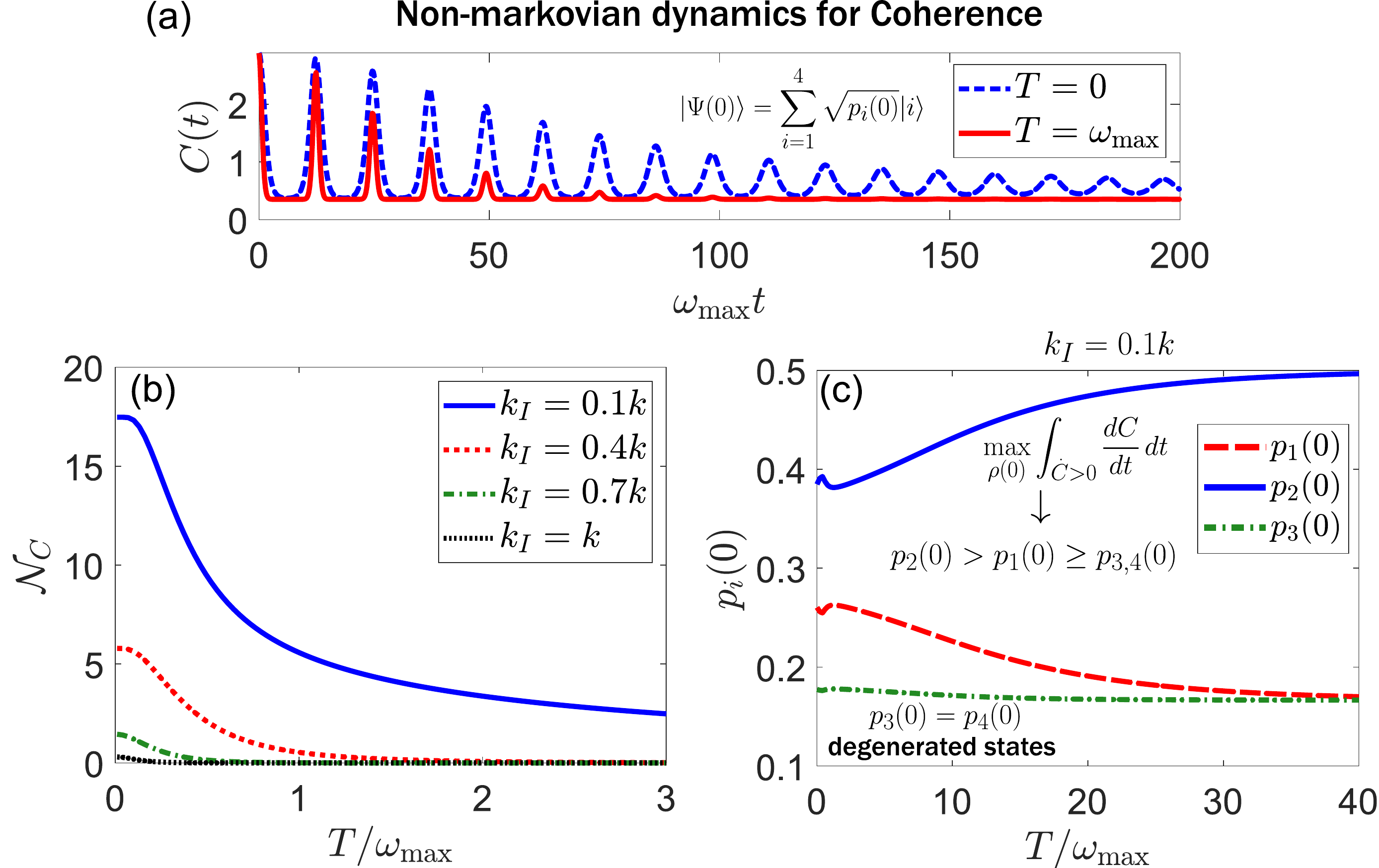}
\caption{(a) Oscillatory behavior of Coherence $C(t)$ indicating a non-Markovian dynamics induced by a pure dephasing channel. The initial state is considered as $|\Psi(0)\rangle = \sum_{i}p_i^{1/2}(0)|i\rangle$, where $p_i(0) = r_i^2$ are the values that solves the optimization problem~(\ref{OptimizationNc}) and $|i\rangle$ are Bell states. (b) Thermal behavior of NM measure $\mathcal{N}_C$ for systems having different elastic constants $k_I$. (c) Optimal initial populations $p_i(0)$ that solve the optimization problem~(\ref{OptimizationNc}) as a function of temperature.}
\label{fig:Figure7}
\end{figure}

In figure~\ref{fig:Figure7}(a) we observe the evolution of the Coherence for temperatures $T=0$ and $T= \omega_{\rm max}$. We considered $m=M/2=1$, $K=k$, and $k_I = 0.1 k$ to simulate strong spin-phonon couplings. We observe that a pure dephasing channel with a narrow coupling can induce oscillations in the Coherence, whose revivals deteriorate if the temperature increases. Therefore, one expects that a non-Markovian measure based on Coherence will capture this thermal effect. \par

It is well-known that Coherence does not increase under incoherent completely positive and trace-preserving (ICPTP) maps if the dynamics is Markovian~\cite{Baumgratz,Chanda}, \textit{i.e.} $dC/dt \leq 0$. Hence, any violation of this monotonicity criteria, given by $dC/dt>0$, can be used to detect non-Markovianity under ICPTP maps. We use the measure of the degree of NM introduced in Ref.~\cite{Chanda}

\begin{equation}\label{NM_measure}
	\mathcal{N}_C = \max_{\rho(0) \in \mathcal{I}^c}\int_{dC(t)/dt>0} {dC(t) \over dt} \, dt, 
\end{equation}
	
where the maximization takes into account initial states $\rho(0)$ belonging to the set of coherent states $\mathcal{I}^{c}$. The above measure of NM is only based on the increasing behavior of the Coherence; therefore, the dynamics is relevant. Let us consider the following set of initial pure states

\begin{equation}
 \rho(0) = |\Psi(0)\rangle \langle \Psi(0)|, \quad \quad |\Psi(0)\rangle = \sum_{j=1}^{4}c_j |j\rangle, 
\end{equation}

where $c_j = r_j e^{i\varphi_j}$ are complex numbers ($r_j \geq 0$ and $0 \leq \varphi_j \leq 2\pi $), and $|j \rangle$ are the eigenstates of the system Hamiltonian described by the Bell states. Then, by using this parametrization of initial conditions, we obtain the following expression for the Coherence

\begin{equation}\label{CoherenciaG}
    C(t)  =  2 \sum_{i<j}\sqrt{p_ip_j} e^{-\xi_{ij}\Upsilon(t)}, 
\end{equation}

where $p_i = r_i^2 = \langle i | \rho(0) |i \rangle$ gives the initial population in each Bell state $|i\rangle$. The measure of NM given in equation~(\ref{NM_measure}) can be written as the following non-linear optimization problem

\begin{eqnarray}
    \mathcal{N}_C &=& \max_{p_1, p_2, p_3, p_4} \mathcal{G}(p_1,p_2,p_3,p_4), \quad 0\leq p_i \leq 1, \quad  \sum_{i=1}^{4}p_i = 1,  \label{OptimizationNc}
\end{eqnarray}

where the non-linear function $\mathcal{G}(\cdot)$ is defined as

\begin{eqnarray}
 \mathcal{G}(p_1,p_2,p_3,p_4) &=& \sum_{i<j} \sqrt{p_ip_j} W_{ij}, \quad  W_{ij} = -2\xi_{ij} \int_{\gamma_c<0} \gamma_c(t) e^{-\xi_{ij} \Upsilon(t)}  \, dt. \label{Mij}
\end{eqnarray}

Because of the symmetry of the matrix elements $\xi_{ij}$ (see Eq.~(\ref{Gamma_xi})), we deduce that $W_{ij} = W_{ji}$ and $W_{ii} = 0$. Also, since the integration presented in equation~(\ref{Mij}) is over the region $\gamma_c(t)<0$, it follows that $W_{ij} \geq 0$. The optimization problem~(\ref{OptimizationNc}) is formally stated in the Supplemental Material. In figure~\ref{fig:Figure7}(b), we show the thermal behavior of $\mathcal{N}_C$ for scenarios where the elastic constant $k_I$ changes. First, we corroborate our previous observations shown in figure~\ref{fig:Figure7}(a); mainly, the NM decreases if temperature increases. Second, for $k_I = 0.1k$, we note that the NM degree increases significantly, showing the effect of the strong and narrow spin-phonon coupling given in figure~\ref{fig:Figure5}. Therefore, to engineer a one-dimensional phonon chain with a large degree of NM in terms of Coherence is necessary to decrease the temperature and have a strong coupling with a particular phonon mode. \par 

Interestingly, this strong coupling is achieved when the effective elastic constant between the lattice and the defect is small. Finally, the optimal initial population $p_i(0) = \rho_{ii}(0)$ that gives us the maximal degree of non-Markovianity, $\mathcal{N_C}$, depends on temperature, as shown in figure~\ref{fig:Figure7}(c). We numerically found that the initial populations always have to fulfill $p_2(0) > p_1(0) \geq p_3(0) = p_4(0)$, where in the high-temperature regime $p_1(0) = p_{3,4}(0)$. In \ref{AppendixD} we present a theoretical calculation showing that $p_3(0) = p_4(0)$ at any temperature. These results imply that the major role in this NM dynamics is hidden in a large initial population of the dark state $|2\rangle$, where $p_{3,4}(0)$ are mostly temperature-independent. Physically, this dephasing channel $L$ given in equation~(\ref{Master_Eq}) does not alter state $|2\rangle$, but in some way, all states move towards the dark state $|2\rangle$, explaining why $p_2(0)$ plays the most relevant contribution to maximize $\mathcal{N}_C$.

\section{Non-markovian measures and physical observables}

\subsection{Canonical rate}

The NM degree based on the canonical rate, $\mathcal{N}_{\gamma}$, can be estimated from experimental measurements. In particular, let us consider the expectation value of the magnetization operator $\mathcal{M}_z = S_{1z}+S_{2z}$. In the eigenstate basis $|i\rangle$, one obtains

\begin{equation}
    \langle \mathcal{M}_z(t) \rangle = 2 \mbox{Re}(\rho_{13}(0)) e^{-2 \Upsilon(t)/3}.
\end{equation}

Using the approximate expression $\gamma_c(t)$ given in equation~(\ref{RateTheo}), we get

\begin{equation}
    \Upsilon(t) \approx  \Upsilon_{\infty} \left[1 - e^{-\Gamma t/2}\left(\cos(\omega_{\rm loc t}) + {\Gamma \over 2 \omega_{\rm loc}} \sin(\omega_{\rm loc} t)\right) \right],
\end{equation}

where $\Upsilon_{\infty} = \lim_{t \rightarrow \infty}\Upsilon(t)$ is the steady-state value, and is given by 

\begin{equation}
    \Upsilon_{\infty}   =  {2 \over \coth\left( {\pi \Gamma  \over 2 \omega_{\rm loc}}\right)} \mathcal{N}_{\gamma},
\end{equation}

with $\mathcal{N}_{\gamma}$ being the NM measure~(\ref{N_gamma}) evaluated at $t \rightarrow \infty$. These expressions allow us to estimate the accumulated degree of NM from $t=0$ to $t = \infty$ using the following relation

\begin{equation}
    \mathcal{N}_{\gamma} \approx {3 \over 2} \coth\left( {\pi \Gamma  \over 2 \omega_{\rm loc}}\right) \mbox{ln}\left({2 \mbox{Re}(\rho_{13}(0)) \over \langle \mathcal{M}_z(\infty) \rangle} \right).
\end{equation}

To implement the above formula, it is required to measure the steady-state state $\langle \mathcal{M}_z(\infty) \rangle$, and also to have some known initial coherence between state $|1\rangle$ and $|3\rangle$. Such a Coherence can be generated by applying a local magnetic field.
 
\subsection{Coherence}
In this section, we present a brief description related to initialization and measurements of Coherence. In particular, if a local and time-dependent magnetic field $\mathbf{B}(t) = (B_x(t), B_y(t), B_z(t))$ is applied on particle $1$ (spin $\mathbf{S}_1$), the bare Hamiltonian of the defect can be written as:

\begin{equation} \label{Hlocal}
      H_{\rm local} = H_{\rm s }  + \gamma \mathbf{B}(t) \cdot \mathbf{S}_1 = \left(\begin{array}{cccc}
    E_1 & -\frac{\gamma}{2}B_{x} & \frac{\gamma}{2}B_z  &-\frac{i\gamma}{2}B_{y}\\
    -\frac{\gamma}{2}B_{x}  & E_{2} &  -\frac{i\gamma}{2}B_{y} & \frac{\gamma}{2}B_{z}\\
    \frac{\gamma}{2}B_{z} & \frac{\gamma}{2}B_{y}  & E_{3} &\frac{\gamma}{2}B_{x}\\
    \frac{i\gamma}{2}B_{y} & \frac{\gamma}{2}B_{z} & \frac{\gamma}{2}B_{x} &  E_{4}\\
    \end{array}\right),
\end{equation}

where $\gamma$ is the gyromagnetic ratio for the system. This result is very useful for quantum control purposes. In particular, one can engineer a control Hamiltonian $H_c(t) = \gamma \mathbf{B}(t) \cdot \mathbf{S}_1$ by using a time-dependent magnetic field acting locally. As we observe in equation~(\ref{Hlocal}), this local field induces all possible transitions between Bell states $|i\rangle$. Therefore, this control Hamiltonian can be used to initialize the system in a pure state $|\Psi(0)\rangle = \sum_i c_i |i\rangle$, which is required to generate some initial Coherence. Recently, a quantum control protocol based on physics-informed neural networks has been proposed to find the unknown optimal control field $B_{\alpha}(t)$ ($\alpha = x,y,z$) in open quantum system~\cite{Norambuena2022}. The computational calculation of the optimal components $(B_{x}(t),B_{y}(t),B_{z}(t))$ is beyond the scope of this work, but a summary of the required dynamical equations is presented in~\ref{AppendixE}. \par 

In addition, we present an more suitable expression for the Coherence in terms of physical observables. We analytically found that

\begin{equation}\label{Cohe}
    C(t)= 2 \sum_{i<j} |\rho_{ij}(t)| = 2\sum_{i=1}^{6}|\langle O_{i}\rangle|,
\end{equation}

where the bilinear operators $O_i$ are defined as $O_1= \mathcal{M}_{z}^{-}\mathcal{M}_{y}^{+}$,  $O_2= \mathcal{M}_{x}^{-}\mathcal{M}_{y}^{-}$, $O_3= \mathcal{M}_{z}^{-}\mathcal{M}_{x}^{-}$, $O_4= \mathcal{M}_{z}^{+}\mathcal{M}_{x}^{-}$, $O_5= \mathcal{M}_{x}^{+}\mathcal{M}_{y}^{-}$, $O_6= \mathcal{M}_{z}^{-}\mathcal{M}_{y}^{-}$. Here, $\mathcal{M}_{\alpha}^{\pm}=S_{2\alpha}\pm S_{1\alpha}$ with $\alpha =x,y,z$. This result tells us that quantum Coherence depends upon the experimental measurement of six physical observables $\langle O_i \rangle$, which correspond to simultaneous measurements on both spins. Therefore, one needs to apply a fast and local control field $\mathbf{B}(t)$ (in a timescale shorter than dissipation), which allows initializing the system in a pure state $|\Psi(0)\rangle = \sum_i c_i |i\rangle$. Then, by measuring the six physical observables $\langle O_{i}\rangle$, one can predict the dynamical behavior of the Coherence.

\section{Conclusions}

In summary, we have presented a detailed analysis of the non-Markovian dynamics arising from defect-phonon interactions in a one-dimensional lattice. We introduced the elastic constant $k_I$ as a key parameter to model how the defect is coupled to the phonon environment. By considering the dipolar magnetic interaction between two spin-$1/2$ particles (defect), we derived the spectral density function (SDF) $J(\omega)$ that describes the behavior of spin-phonon coupling. We deduced the time-local master equation for the reduced density matrix of the defect (two spins), obtaining a pure-dephasing channel acting on Bell states. Also, the role of the energy spectrum and the pure-dephasing channel are physically discussed. \par 

The non-Markovian dynamics of the defect is analyzed in terms of two independent measures ($\mathcal{N}_{\gamma},\mathcal{N}_{C}$) based on: the canonical rate $\gamma_{c}(t)$ and the Coherence $C(t)$. First, we found an analytical representation for the canonical rate that explains all features of $\mathcal{N}_{\gamma}$, in particular, the role of temperature, width $\Gamma$ and phonon frequency $\omega_{\rm loc}$. These findings allow us to connect how the elastic constant $k_I$ and temperature $T$ can modify the behavior of the rate and NM measure $\mathcal{N}_{\gamma}$. Second, by finding a generic analytical solution for the reduced density matrix for the defect, we formally wrote the non-linear optimization problem to compute the degree of NM based on the Coherence, $\mathcal{N}_C$. Here, we noted that temperature deteriorates the degree of NM $\mathcal{N}_C$, and only strong-phonon coupling characterized by a narrow SDF $J(\omega)$ can increase the NM behavior. Moreover, for optimizing this short-memory effect it is required that $p_2(0) \geq p_1(0) = p_3(0) = p_4(0)$, where $p_i(0)$ are the initial populations for a pure state $|\Psi(0)\rangle = p_i^{1/2}(0)|i\rangle$, where $|i\rangle$ are Bell states. This finding shows that for a four-level system with a pure-dephasing channel, $L = -\sqrt{2/3}[\ket{1}\bra{1}-{1 \over 2}(\ket{3}\bra{3}+\ket{4}\bra{4} )]$, the dark state $|2\rangle$ plays a fundamental role in maximizing the degree of NM. Furthermore, we provided a recipe to implement these ideas through physical observables by applying global and local magnetic fields, as well as a more experimental accessible expression for the Coherence. This work showcase a deep understanding upon the origin of non-Markovianity, and delivers ways for engineering it in real-world systems such as phononic crystals with molecular defects with residual spin-$1/2$. 

\appendix

\section{Role of local normal modes of the spin defect} \label{AppendixA}

\begin{figure}[ht]
\centering
\includegraphics[width=0.8 \linewidth]{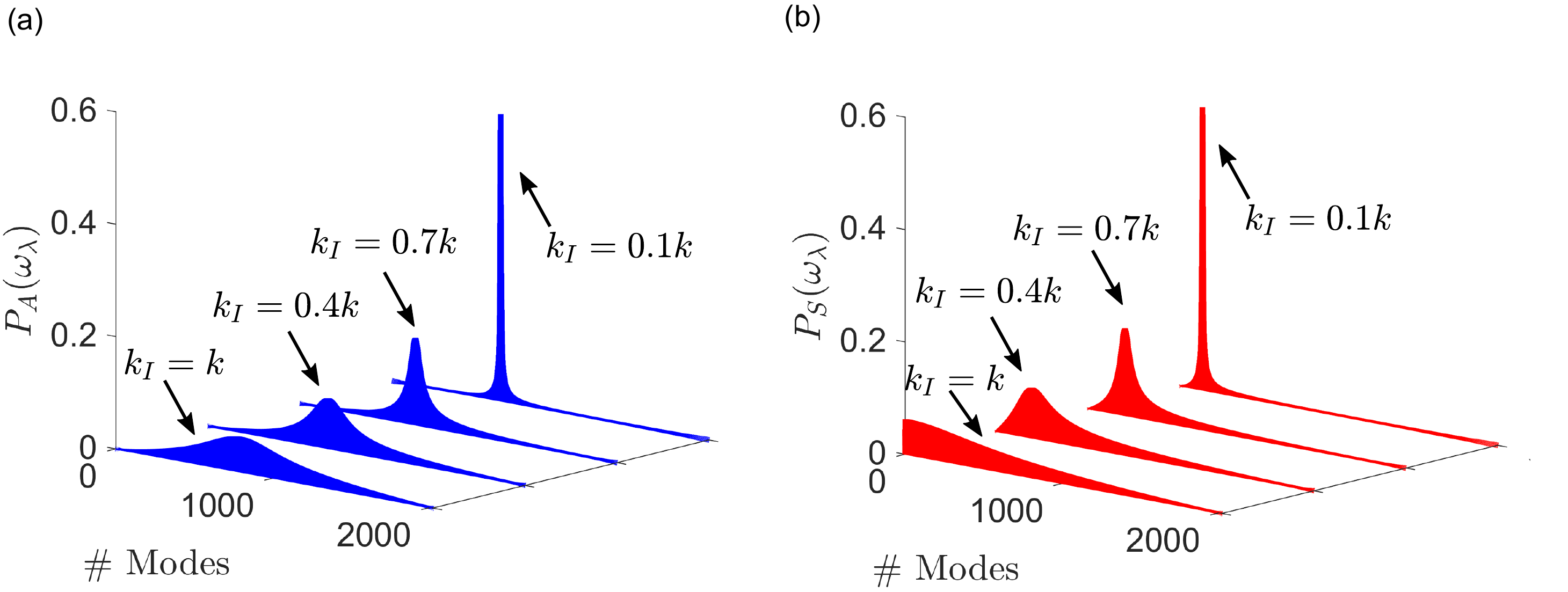}
\caption{Projection of the system modes onto the antisymmetric (a) and symmetric (b) local modes of the dimer defect.}
\label{fig:Figure2}
\end{figure} 

Further insights into the vibrational properties of the defect-phonon system can be gained if the projection of defect normal coordinates $Q_A$ and $Q_S$, see equation~(\ref{LocalNormalModes}) are projected 
into the normal modes of the full system (defect + phonon chain). To this end, we introduce the projections:

\begin{eqnarray} \label{Projection}
P_{\lambda'}(\omega_{\lambda}) = \mathbf{h}_{\lambda}\cdot \mathbf{h}_{\lambda'}^{\rm defect}, \quad  \lambda' = S,A.
\end{eqnarray}

First, $\mathbf{h}_{\lambda} = \sum_{i}h_{\lambda i}\mathbf{e}_i$ is a normalized real vector defined in terms of the eigenvectors $h_{\lambda i}$ that are solutions of the eigenvalue equation~(\ref{EigenvalueEquationDij}). Second, $\mathbf{h}_{\lambda'}^{\rm defect} = \sum_{i}h_{\lambda i}^{\rm defect}\mathbf{e}_i$ is another normalized real vector for the local symmetric ($\lambda = S$) and antisymmetric ($\lambda = A$) normal modes of the spin defect. For instance, in our case we have $h^{\rm defect}_{S,i}=[0,...,0,1,1,0,...0]^T/\sqrt{2}$ and $h^{\rm defect}_{A,i}= [0,...,0,-1,1,0,...0]^T/\sqrt{2}$. In Fig.~\ref{fig:Figure2}(a)-(b) we calculate the projections $P_{S}(\omega_{\lambda})$ and $P_{A}(\omega_{\lambda})$ for different internal elastic constants $k_I$ using $N=2022$, $K=k=1$, and $M=2m=2$. It is worth noting that for small values of the defect-lattice coupling $k_I$, the projections $P_{\lambda'}(\omega_{\lambda})$ become narrow and sharper, suggesting that dimer's local modes are activated at particular frequencies for $\lambda' = S,A$. Note that our defect is represented by two masses attached with an internal spring constant $K$. Therefore, the defect has a characteristic breathing mode with frequency $\omega_{\rm breathing} = \sqrt{2 K /M} \approx 1$, which is very close to the peak of $P_{A}(\omega_{\lambda})$ in the limit $k_I \ll k$, see Fig.~\ref{fig:Figure2}(b). The latter suggests the emergence of a strong vibrational resonance of the dimer defect in the regime $k_I \ll k$. 

\section{Microscopic derivation of the time-local master equation} \label{AppendixB}

The dynamics of the system is given by the von Neumann equation $(\hbar = 1)$

\begin{equation}\label{Eq_von_Neumann}
  \frac{d\rho_{\rm T}(t)}{dt}=-i[H, \rho_{\rm T}(t)]
\end{equation}

where $\rho_{\rm T}$ is the defect-phonon density operator, $H$ is the defect-phonon Hamiltonian and is written as

\begin{equation}\label{total_hamiltonian}
  H = H_{\rm s}+H_{\rm ph}+H_{\rm s-ph}
\end{equation}

with $H_{\rm s}=C/r^{3} F(\mathbf{S}_1,\mathbf{S}_2)$ is the defect Hamiltonian, $H_{\rm ph}=\sum_{\lambda}\omega_{\lambda}b^{\dagger}_{\lambda}b_{\lambda}$, is the phonon Hamiltonian, and the defect-phonon interaction

\begin{equation}\label{Inter-hamil}
  H_{\rm s-ph} = \sum_{\lambda} g_{\lambda} (b_{\lambda} + b_{\lambda}^{\dagger} ) F(\mathbf{S}_1,\mathbf{S}_2),
\end{equation}

with $F(\mathbf{S}_1,\mathbf{S}_2) = 2 S_{1x}S_{2x}- S_{1y}S_{2y}-S_{1z}S_{2z}$. In equation~(\ref{Inter-hamil}), we observe that only one dissipation channel corresponds to a pure dephasing channel. We move to the interaction picture and formally integrating equation~(\ref{Eq_von_Neumann}), we obtain the defect-phonon density operator in the interaction picture

\begin{equation}\label{sol_eq}
  \tilde{\rho}_{\rm T}(t)= \tilde{\rho}_{\rm T}(0)-i\int_{0}^{t}dt'[\tilde{H}_{\rm s-ph}(t'),\tilde{\rho}_{\rm T}(t')],
\end{equation}

where $\tilde{\rho}_{\rm T}(t)=e^{iH_{0}t}\rho_{\rm T}e^{-iH_{0}t}$ with $H_{0}=H_{\rm s}+H_{\rm ph}$. Substituting the equation~(\ref{sol_eq}) into the equation~(\ref{Eq_von_Neumann}) and calculating the partial trace over the phonon degrees of freedom, we obtain the following equation for the defect density matrix $\rho_{\rm s}$ ($\hbar = 1 $)

\begin{eqnarray}\label{rho_s}
  {d\rho(t) \over dt}&=&-i\hbox{Tr}_{\rm ph}\left([\tilde{H}_{\rm s-ph}(t),\tilde{\rho}_{\rm T}(0)]\right)\nonumber\\
  &&\quad-\int_{0}^{t}dt'\hbox{Tr}_{\rm ph}\left([\tilde{H}_{\rm s-ph}(t),[\tilde{H}_{\rm s-ph}(t'),\tilde{\rho}_{\rm T}(t')]]\right),
\end{eqnarray}

where $\tilde{\rho}_{s}(t)=\rho(t)=\hbox{Tr}_{\rm ph}(\tilde{\rho}_{\rm s-ph})$. We assume that at any time, the full density operator can decompose as a product state $\tilde{\rho}_{\rm s-ph}=\tilde{\rho}_{\rm s}\otimes\rho_{\rm ph}$ because we consider the weak coupling limit $g_{\lambda}\ll |C/r^{3}|$. For a thermal phonon state we have $\hbox{Tr}_{\rm ph}[\hat{b}_{\lambda}\hat{\rho}_{\rm ph}]=\hbox{Tr}_{\rm ph}[\hat{b}_{\lambda}^{\dagger}\hat{\rho}_{\rm ph}]=0$, then $\hbox{Tr}_{\rm ph}\left([\tilde{H}_{\rm s-ph}(t),\tilde{\rho}_{\rm T}(0)]\right)=0$. In thermal equilibrium, the phonon density operator is given by $\rho_{\rm ph}\exp{(-\beta \hat{H}_{b}/Z)}$, where $Z=\hbox{Tr}_{\rm ph}[\exp{(-\beta \hat{H}_{\rm ph})}]$ is the partition function and $\beta = (k_{B}T)^{-1}$. Thus, the equation~(\ref{rho_s}) read

\begin{equation}\label{rho_s2}
  {d\rho(t) \over dt}=-\int_{0}^{t}dt'\hbox{Tr}_{\rm ph}\left([\tilde{H}_{\rm s-ph}(t),[\tilde{H}_{\rm s-ph}(t'),\tilde{\rho}_{\rm T}(t')]]\right).
\end{equation}

We write the interaction Hamiltonian as $\tilde{H}_{\rm s-ph}=(G(t)+G^{\dagger}(t))\tilde{F}(t)$ with $G(t)=\sum_{\lambda}g_{\lambda}\hat{b}_{\lambda}e^{-i\omega_{\lambda}t}$ and $G^{\dagger}(t)=\sum_{\lambda}g_{\lambda}\hat{b}^{\dag}_{\lambda}e^{i\omega_{\lambda}t}$. Developing the commutator of equation~(\ref{rho_s2}), applying the first and second Markov approximations, and assuming that $\rho(t')\approx \rho(t)$, we obtain for $t\longrightarrow \infty$~\cite{deVega2017, Breuer2010,Kryszewski}

 \begin{eqnarray}\label{Master_Eq}
  \frac{d\rho(t)}{dt}&=&\int_{0}^{\infty}d\tau\left(A_{\lambda}(\tau)+B_{\lambda}^{\ast}(\tau)\right)\times\nonumber\\
  &&\quad\quad\left(\tilde{F}(t-\tau)\rho(t)\tilde{F}^{\dag}(t)-\tilde{F}^{\dag}(t)\tilde{F}(t-\tau)\rho(t)\right) + h.c.,
\end{eqnarray}

where $A_{\lambda}(\tau)=\hbox{Tr}_{\rm ph}[G^{\dagger}(t)G(t')\rho_{\rm ph}]=\langle G^{\dagger}(t)G(t')\rangle=\sum_{\lambda}|g_{\lambda}|^{2}n(\omega_{\lambda})e^{i\omega_{\lambda}\tau}$, $B_{\lambda}(\tau)=\hbox{Tr}[G(t)G^{\dagger}(t')\rho_{\rm ph}]=\langle G(t)G^{\dagger}(t')\rangle=\sum_{\lambda}|g_{\lambda}|^{2}[n(\omega_{\lambda})+1]e^{-i\omega_{\lambda}\tau}$ and $\tau = t-t'$. Here, $n(\omega_{\lambda})=[\exp(\hbar\omega_{\lambda}/k_{B}T)-1]^{-1}$ is the mean number of phonons at thermal equilibrium. Now, we introduce the spectral decomposition~\cite{Kryszewski},

\begin{equation}
  F(\mathbf{S}_1,\mathbf{S}_1)=\sum_{\omega}\hat{F}(\omega),
\end{equation}

\begin{equation}
  \hat{F}(\omega)=\sum_{a,b}\delta(\omega_{ba}-\omega)|a\rangle\langle a|F(\mathbf{S}_1,\mathbf{S}_1)|b\rangle\langle b|,
\end{equation}

where $\delta(\omega_{ba}-\omega)$ is a kronecker function. The quantum states $\ket{a}$, $\ket{b}$ are eigenstates of the Hamiltonian $H_{\rm s}$. In the interaction picture, the following relations are satisfied in the frequency domain

\begin{equation}\label{Ft}
  \hat{F}(t)=\sum_{\omega}e^{-i\omega t}\hat{F}(\omega), \quad \hat{F}(t-\tau)=\sum_{\omega'}e^{-i\omega'(t-\tau)}\hat{F}(\omega')
\end{equation}

Substituting the equation~(\ref{Ft}) into equation~(\ref{Master_Eq}), we will obtain oscillating terms proportional to $\exp(\pm i(\omega'-\omega)t)$. In the secular approximation, we neglect the terms $\omega\neq \omega'$ due to the condition $\tau_{\rm ph}\gg T_{\rm s}$, where $\tau_{\rm ph}$ and $T_{s}$ are the bath and system characteristic times, respectively. Thus, we obtain
\begin{eqnarray}
      {d\rho(t) \over dt}&=&\sum_{\omega}\Gamma(\omega,t)\left(\tilde{F}(\omega)\rho(t)\tilde{F}^{\dag}(\omega)-\tilde{F}^{\dag}(\omega)\tilde{F}(\omega)\rho(t)\right)\nonumber\\
  &&+\sum_{\omega}\Gamma^{\ast}(\omega,t)\left(\tilde{F}(\omega)\rho(t)\tilde{F}^{\dag}(\omega)-\rho(t)\tilde{F}^{\dag}(\omega)\tilde{F}(\omega)\right),
\end{eqnarray}

where

\begin{equation}
  \Gamma(\omega,t)=\int_{0}^{\infty}d\tau|g_{\lambda}|^{2}\left(n(\omega_{\lambda})e^{i\omega_{\lambda}\tau}+[n(\omega_{\lambda})+1]e^{-i\omega_{\lambda}\tau}\right)e^{i\omega \tau},
\end{equation}

with $\Gamma^{\ast}(\omega,t)$ being the complex conjugate of $\Gamma(\omega,t)$. Now, we define
\begin{eqnarray}
  \gamma(\omega,t)&=&\Gamma(\omega,t)+\Gamma^{\ast}(\omega,t),\\
  \lambda(\omega,t)& =&\frac{1}{2i}(\Gamma(\omega,t)-\Gamma^{\ast}(\omega,t)).
\end{eqnarray}  

By using $\gamma(\omega,t)$ and $\lambda(\omega,t)$, we obtain recover the time-local master equation given in equation~(\ref{TimeLocalMasterEquation}), where the Lamp-shift Hamiltonian has been neglected because of the weak-coupling approximation (which was numerically corroborated).

\section{Approximated expressions for canonical rate and degree of non-Markovianity $\mathcal{N}_{\gamma}$}~\label{AppendixC}

Using the phenomenological spectral density function defined in equation~(\ref{SDF}) into the definition of the canonical rate, see equation~$\gamma_c(t)$, we obtain 

\begin{equation}
    \gamma_c(t) = 3 J_0 \int_{0}^{\infty}{1 \over \omega} \sin^{1+n}\left({\omega \over \omega_{\rm loc}}\pi \right)  L(\omega-\omega_{\rm loc}) \mbox{coth}\left( {\hbar \omega \over 2 k_B T} \right)\sin(\omega t) \, d\omega, \label{Integral1}
\end{equation}

where $L(x)$ is the Lorentzian function centered around $x = 0$

\begin{equation}
  L(x)  =   {(\Gamma/2) \over x^2+ (\Gamma/2)^2}.
\end{equation}

Because of the $L(\cdot)$ function, one observes that the main contribution to the integral~(\ref{Integral1}) comes from the frequencies around $\omega = \omega_{\rm loc}$. We apply the change of variable $u = \omega-\omega_{\rm loc}$, and we extend the integration region for $u \in (-\infty, \infty)$ (since $L(x)$ is zero for regions beyond $x=0$), obtaining

\begin{equation}
    \gamma_c(t) = 3 J_0 \int_{-\infty}^{\infty} F(u) L(u)  \, du,
\end{equation}

where the function $F(u)$ is defined as

\begin{eqnarray}
    F(u) &=& g(u)\sin(ut)\cos(\omega_{\rm loc}t)  + g(u)\cos(ut)\sin(\omega_{\rm loc}t), \\
    g(u) &=& {1\over \omega}\sin^{1+n}\left({(u+\omega_{\rm loc}) \over \omega_{\rm loc}}\pi \right)  \mbox{coth}\left( {\hbar (u+\omega_{\rm loc}) \over 2 k_B T} \right),
\end{eqnarray}

where we have used $\sin(x+y) = \sin(x)\cos(y)+\cos(x)\sin(y)$. Now, we approximate the function $g(u)$ as $g(0)$ since the $L(u)$ takes the major contribution around $u=0$. We get
\begin{eqnarray}
    \gamma_c(t) &\approx& 3 J_0 g(0) \cos(\omega_{\rm loc}t)\int_{-\infty}^{\infty} \sin(ut) L(u)  \, d\omega \nonumber \\                       &       & +3 J_0 g(0) \sin(\omega_{\rm loc}t)\int_{-\infty}^{\infty} \cos(ut) L(u)  \, d\omega.
\end{eqnarray}

By symmetry considerations we deduce that $\int_{-\infty}^{\infty} \sin(ut) L(u)  \, d\omega  = 0$, and the other integral can be analytically solved as

\begin{equation}
    \int_{-\infty}^{\infty} \cos(ut) L(u)  \, d\omega =  \pi e^{-\Gamma t/2}.
\end{equation}

We recover the approximated expression given in equation~(\ref{RateTheo}) using these previous calculations. The degree of non-Markovianity $\mathcal{N}_{\gamma}$ is calculated using $\gamma_c(t) \approx \Lambda(T) \sin(\omega_{\rm loc}t)e^{-\Gamma t/2}$, which gives

\begin{eqnarray}
    \mathcal{N}_{\gamma} &= & {1 \over 2}\int_{0}^{\infty} \left(\vert\gamma_c (\tau) \vert -\gamma_c (\tau)\right)\, d\tau \nonumber \\
    & \approx & {1 \over 2} \Lambda(T) \int_{0}^{\infty} e^{-\Gamma \tau/2}\left(|\sin(\omega_{\rm loc}\tau)|-\sin(\omega_{\rm loc}\tau) \right) \, d \tau.
\end{eqnarray}

The function $h(\tau) = |\sin(\omega_{\rm loc}\tau)|-\sin(\omega_{\rm loc}\tau)$ is non-zero only in the time intervals $\tau \in [(2n+1)\pi/\omega_{\rm loc},(2n+2)\pi/\omega_{\rm loc}]$ for $n  \in \mathds{N}_0$. In such intervals, the function assume the value $h(\tau) = 2\sin(\omega_{\rm loc}\tau)$. Therefore, we can use the following expression

\begin{eqnarray}
    \mathcal{N}_{\gamma}  &\approx&  \Lambda(T) \sum_{n=0}^{\infty}\int_{(2n+1)\pi/\omega_{\rm loc}}^{(2n+2)\pi/\omega_{\rm loc}} e^{-\Gamma \tau/2}\sin(\omega_{\rm loc}\tau)\, d \tau \nonumber \\
    &=&  {1 \over 2}\Lambda(T) {\omega_{\rm loc} \over \omega_{\rm loc}^2+(\Gamma/2)^2}\left(e^{\beta}+1 \right)\sum_{n=0}^{\infty} e^{-\beta(n+1)},
\end{eqnarray}

where $\beta = \pi \Gamma/(2\omega_{\rm loc})>0$. By solving the geometric series $\sum_{n=0}^{\infty} e^{-\beta(n+1)} = [e^{\beta}-1]^{-1}$, we obtain the equation~(\ref{NM_theo}). 

\section{Optimization problem for the degree of non-Markovianity $\mathcal{N}_C$} \label{AppendixD}

The target function of the optimization problem presented in equation~(\ref{OptimizationNc}) can be written as

\begin{eqnarray}
  \mathcal{G}(r_1,r_2,r_3,r_4) = W_1\left(r_1r_3 + r_1 r_4 \right)  +  W_2\left(r_1r_2 + r_2 r_3 + r_3 r_4 \right),
\end{eqnarray}

where $W_1 \equiv W_{13} = W_{14}$, $W_2 \equiv W_{12} = W_{23} = W_{24}$, and $r_i = \sqrt{p_i}$ ($i =1,2,3,4$). To solve the optimization problem, we need to solve the system of equations $\partial \mathcal{G} / \partial r_i =0$ for $i = 1,2,3,4$. In particular, when one impose $\partial \mathcal{G} / \partial r_3 =0$, we obtain the equation

\begin{equation} 
 {\partial \mathcal{G} \over \partial r_3} = 0 \quad \longrightarrow \quad \left(W_1 r_1 + W_2 r_2 \right)\left(1-{r_3 \over \sqrt{1-r_1^2-r_2^2-r_3^2}} \right)  = 0,
\end{equation}

where we have used the probability constraint $\sum_i r_i^2 = 1$ to write $r_4 = (1-r_1^2-r_2^2-r_3^2)^{1/2}$. Since $W_1>0$, $W_2>0$ and $r_i \geq 0$, we deduce that

\begin{equation}
    1={r_3 \over \sqrt{1-r_1^2-r_2^2-r_3^2} }\quad \longrightarrow \quad r_1^2 + r_2^2+2r_3^2 = 1.
\end{equation}

The above condition and the constraint $\sum_i r_i^2 = 1$ imply that $r_3 = r_4$. Therefore, to optimize the target function $\mathcal{G}(\cdot)$, it is required that the initial condition satisfy $p_3(0) = p_4(0)$, which is a consequence of the degeneracy of states $|3\rangle$ and $|4\rangle$.

\section{Hamiltonian with global and local magnetic field} \label{AppendixE}

In this appendix, we consider the closed dynamics of the spin defect under the effect of a global and local magnetic field. This is a good approximation when the control Hamiltonian acts in a timescale shorter than the phonon-induced dissipation time. Let us consider a magnetic field $\mathbf{B}=(B_{x},B_{y},B_{z})$ such that the Hamiltonian of the system reads as,
\begin{equation}
  H= \sum_{i=1}^{4}E_{i}\ket{i}\bra{i}+B_{x}(S_{1x}+S_{2x})+B_{y}(S_{1y}+S_{2y})+B_{z}(S_{1z}+S_{2z})
\end{equation}

The operators $S_{i\alpha}$ with $\alpha=x,y,z$ in the base of the eigenstates~(\ref{e1})-(\ref{e4}) are given by

\begin{eqnarray}
  S_{1x}&=&\frac{1}{2}(-\ket{1}\bra{2}-\ket{2}\bra{1}+\ket{3}\bra{4}+\ket{4}\bra{3})\label{Sx1}\\
  S_{2x}&=&\frac{1}{2}(\ket{1}\bra{2}+\ket{2}\bra{1}+\ket{3}\bra{4}+\ket{4}\bra{3})\label{Sx2}\\
  S_{1y}&=&\frac{i}{2}(-\ket{1}\bra{4}+\ket{4}\bra{1}-\ket{2}\bra{3}+\ket{3}\bra{2})\label{Sy1}\\
  S_{2y}&=&\frac{i}{2}(-\ket{1}\bra{4}+\ket{4}\bra{1}+\ket{2}\bra{3}-\ket{3}\bra{2})\label{Sy2}\\
  S_{1z}&=&\frac{1}{2}(\ket{1}\bra{3}+\ket{3}\bra{1}+\ket{2}\bra{4}+\ket{4}\bra{2})\label{Sz1}\\
  S_{2z}&=&\frac{1}{2}(\ket{1}\bra{3}+\ket{3}\bra{1}-\ket{2}\bra{4}-\ket{4}\bra{2})\label{Sz2}
\end{eqnarray}

Then, the Hamiltonian can be written as

\begin{equation}\label{Ham_1}
  H = \sum_{i=1}^{4}E_{i}\ket{i}\bra{i}+B_{x}(\ket{3}\bra{4}+h.c.)+B_{y}(-i\ket{1}\bra{4}+h.c.)+B_{z}(\ket{1}\bra{2}+h.c.)
\end{equation}

where $E_{i}$ and $\ket{i}$ are the eigenenergies and eigenstates of the spin defect without a magnetic field. The Hamiltonian~(\ref{Ham_1}) is written in matrix form as
\begin{equation}
  H_{\rm global}=\left(\begin{array}{cccc}
    E_{1} & 0 & B_{z}  &-iB_{y}\\
    0  & E_{2} &  0  & 0\\
    B_{z} & 0  & E_{3} &B_{x}\\
    iB_{y} & 0 & B_{x} &  E_{4}\\
    \end{array}\right)
\end{equation}
When a global field is applied over the two spin-$1/2$ particles, we notice some forbidden transitions. Now, if we apply a local magnetic field, that is, it interacts with one of the spins, let's say the spin $\mathbf{S}_{1}$, the Hamiltonian is written as

\begin{equation}
  H_{\rm local} = \sum_{i=1}^{4}E_{i}\ket{i}\bra{i}+B_{x}S_{1x}+B_{y}S_{1y}+B_{z}S_{1z},
\end{equation}

in matrix form

\begin{equation}
  H_{\rm local}=\left(\begin{array}{cccc}
    E_1 & -\frac{1}{2}B_{x} & \frac{1}{2}B_z  &-\frac{i}{2}B_{y}\\
    -\frac{1}{2}B_{x}  & E_{2} &  -\frac{i}{2}B_{y} & \frac{1}{2}B_{z}\\
    \frac{1}{2}B_{z} & \frac{1}{2}B_{y}  & E_{3} &\frac{1}{2}B_{x}\\
    \frac{i}{2}B_{y} & \frac{1}{2}B_{z} & \frac{1}{2}B_{x} &  E_{4}\\
    \end{array}\right)
\end{equation}

We note that now the Zeeman effect generates all possible transitions between the eigenstates $|i\rangle$. We can use $H_{\rm local}$ to initialize the system to a particular linear combination $|\Psi(0)\rangle = \sum_i c_i |i\rangle$. The latter can be done in a physics-informed neural network that was recently proposed in Ref.~\cite{Norambuena2022}, where $\mathbf{B}(t)$ is a control magnetic field. With this neural network, fast control fields can be obtained that can prepare the system in one of the eigenstates in a short time. The differential equations to implement in the network are obtained from the Schrödinger equation given by,

\begin{equation}
    {\partial \ket{\Psi(t)}\over \partial t}=-{i\over \hbar}  H\ket{\Psi(t)}
\end{equation}

Choosing as ansatz the state vector,
\begin{equation}
  \ket{\Psi(t)}=\sum_{i=1}^{4}c_{i}(t)\ket{i},
\end{equation}

where $c_{i}(t)$ are complex, whose dynamical equations are given by,
\begin{eqnarray}
  {d c_{1}\over dt} &=& -{i\over \hbar}\left(E_{1}c_{1}(t)-{1\over 2}B_{x}c_{2}(t)-{i\over 2}B_{y}c_{4}(t)+{1\over 2}B_{z}c_{3}(t)\right)\\
  {d c_{2}\over dt} &=& -{i\over \hbar}\left(E_{2}c_{2}(t)-{1\over 2}B_{x}c_{1}(t)-{i\over 2}B_{y}c_{3}(t)+{1\over 2}B_{z}c_{4}(t)\right)\\
  {d c_{3}\over dt} &=& -{i\over \hbar}\left(E_{3}c_{3}(t)+{1\over 2}B_{x}c_{4}(t)+{i\over 2}B_{y}c_{2}(t)+{1\over 2}B_{z}c_{1}(t)\right)\\
  {d c_{4}\over dt} &=& -{i\over \hbar}\left(E_{4}c_{4}(t)+{1\over 2}B_{x}c_{3}(t)+{i\over 2}B_{y}c_{1}(t)+{1\over 2}B_{z}c_{2}(t)\right)
\end{eqnarray}

Since the neural network implements only real numbers, we parameterize the coefficients as $c_{j}(t)= c_j^{\rm R}(t) + ic_j^{\rm I}(t)$, where $c_j^{\rm R}(t )= \mbox{Re}(c_j)$ and $c_j^{\rm I}(t) = \mbox{Im}(c_j)$. While the initial condition is parameterized as follows

\begin{equation}
  \ket{\Psi(0)}=\sum_{i=1}^{4}c_{i}(0)\ket{i}
\end{equation}

Finally, we take advantage of the equations~(\ref{Sx1})-(\ref{Sz2}) of the spin operators to write the Coherence in terms of observable ($\langle O\rangle$), and it is given by,

\begin{equation}\label{Cohe2}
    C(t)=2\sum_{i=1}^{6}|\langle O_{i}\rangle|,
\end{equation}
 with $O_1= \mathcal{M}_{z}^{-}\mathcal{M}_{y}^{+}$,  $O_2= \mathcal{M}_{x}^{-}\mathcal{M}_{y}^{-}$, $O_3= \mathcal{M}_{z}^{-}\mathcal{M}_{x}^{-}$, $O_4= \mathcal{M}_{z}^{+}\mathcal{M}_{x}^{-}$, $O_5= \mathcal{M}_{x}^{+}\mathcal{M}_{y}^{-}$, $O_6= \mathcal{M}_{z}^{-}\mathcal{M}_{y}^{-}$ and $\mathcal{M}_{\alpha}^{\pm}=S_{2\alpha}\pm S_{1\alpha}$, $\alpha =x,y,z$. This result is an alternative expression to the coherence~(\ref{CoherenciaG}) presented in the main text. In general, the expression~(\ref{Cohe2}) can be used to obtain the Coherence in our particular four-level system.
 
\section*{References}

% References in JPN style
\bibliographystyle{unsrt}

\end{document}